\pdfoutput=1
\documentclass[12pt, hyper]{article}
\usepackage{amsmath,amsopn,amssymb,amsfonts}
\usepackage{amsthm}
\usepackage{epsfig}
\usepackage{verbatim}
\usepackage{multirow}
\usepackage{arydshln}
\usepackage{cite}
\usepackage{float}
\usepackage{booktabs}
\usepackage{graphicx} % Required for inserting images
\usepackage{amsthm}
\usepackage[margin=1cm]{caption}

\usepackage{hyperref}
\usepackage{nicematrix}           % continuous vertical rules unbroken by \rowcolor
\NiceMatrixOptions{custom-line = {letter=I, color=black!30}} 

\def\b{\beta}

\def\t{\tau}

\usepackage{xcolor}
\usepackage{colortbl}
\definecolor{sph}{HTML}{C72921}   % sphaleron (red), matches plot palette
\definecolor{bnc}{HTML}{2257A8}   % bounce (blue), matches plot palette

\usepackage{cleveref}

\begin{document}

\begin{titlepage}
\vfill
\begin{flushright}
{\tt\normalsize KIAS-P26035}\\
%{\tt\normalsize hep-th/yymmnnn}\\
\end{flushright}
\vfill
\begin{center}
{\Large\bf Resurgence of the Thermal Transition between Bounce and Sphaleron}
%Resurgent Analysis of Bounce-Sphaleron Phase Transition

\vfill

Shaun D. Hampton, Kyungsun Lee, and Sungjay Lee

\vskip 5mm
{\it Korea Institute for Advanced Study
\\85 Hoegiro, Dongdaemun-Gu, Seoul 02455, Korea }

\end{center}
\vfill

\begin{abstract}
\noindent 
We study the thermal transition between the bounce and the sphaleron in quantum mechanics 
with a metastable vacuum from the viewpoint of Borel resurgence. 
For two models representing a second-order and a first-order transition, we compute the perturbative expansion of the thermal free energy to high orders and extract the leading Borel singularity data $(A,b,S)$ as functions of temperature. The Borel singularity location $A$ reproduces the on-shell action of the dominant saddle on both sides of the transition, joining smoothly in the second-order case and developing a kink in the first-order case. The characteristic exponent $b$ jumps between $0$ and $1/2$ across the transition, counting the zero modes of the corresponding saddle.  The Stokes constant $S$ matches the one-loop determinant around the saddle. The perturbative expansion around the false vacuum thus determines the transition temperature, the order of the transition, and the decay rate including the one-loop prefactor without relying on semiclassical inputs.
\end{abstract}

\vfill
\end{titlepage}

\parskip 0.1 cm

\newpage

\renewcommand{\thefootnote}{\#\arabic{footnote}}
\setcounter{footnote}{0}

\parskip 0.2 cm

%======================================================================
\section{Introduction and Conclusion}
\label{sec:intro}
%======================================================================
 
In quantum mechanics with a metastable vacuum, the false-vacuum
decay rate at finite temperature $T = 1/\beta$ can be described by
non-perturbative saddle points of the Euclidean path integral on 
a circle of circumference $\beta$ in the semiclassical limit. 
Here two types of saddles compete.  One of them is a
non-trivial periodic classical solution with period~$\beta$ and
finite Euclidean action.  The other is the static
solution sitting at the barrier top with Euclidean action
$\beta V_{\mathrm{top}}$.  The former is known as the \emph{bounce}
(or periodic instanton), while the latter as the \emph{sphaleron}~\cite{Manton:1983nd, Klinkhamer:1984di}.
At low temperature the bounce dominates, describing quantum
tunneling through the barrier.
At high temperature the sphaleron takes over, corresponding to
classical thermal fluctuation over the barrier top.  

The local shape of the potential near the barrier top determines whether the switch of the dominant saddle from one to the other is smooth or sharp~\cite{Chudnovsky:1992xrj}. If the period of the
bounce, which can be identified as the inverse temperature $\beta$, decreases monotonically
as the energy increases towards the barrier height, then the transition is
continuous and the bounce action joins the sphaleron action smoothly. 
In other words, the transition is second order.  
If instead the period
map develops a fold, meaning it first decreases and then turns back,
then two distinct bounce solutions coexist in an intermediate
temperature range and the transition becomes discontinuous. It describes 
a first-order transition. 
The semiclassical theory of these thermal transitions
was developed in the seminal works of
Langer~\cite{Langer}, Affleck~\cite{Affleck:1980ac}, and
others~\cite{Callan:1977pt, Coleman:1977py, Linde:1981zj}.
 
The thermal transition between the bounce and the sphaleron plays
a central role in several problems of particle physics and
cosmology.  In the electroweak phase transition, the sphaleron rate
governs baryon number violation at high temperatures, while the
bounce controls the nucleation of bubbles of the broken phase at
lower
temperatures~\cite{Kuzmin:1985mm, Arnold:1987mh, Carson:1990jm}.
An accurate description of the transition region is essential for
determining the baryon asymmetry of the
universe~\cite{Anderson:1991zb, Morrissey:2012db}.  Similar issues arise in the
study of vacuum stability, where the thermal transition rate between
the electroweak vacuum and a deeper true vacuum depends sensitively
on which saddle dominates at a given
temperature. In all these applications,
the semiclassical approximation is least reliable precisely near the
transition temperature, where the two saddle-point actions are comparable and
higher-order corrections become important.
 
In a seemingly unrelated line of development, the theory of
resurgence has revealed a deep connection between perturbative and
non-perturbative physics.  It has long been known that perturbative
series in quantum mechanics are generically divergent, with
fast-growing coefficients~\cite{Bender:1969si, Bender:1973rz, Lipatov:1976ny}.
The crucial observation by Bender and Wu~\cite{Bender:1973rz} 
is that this divergence is not a flaw 
but reflects crucial non-perturbative physics.  The rate at which the
coefficients grow encodes the Euclidean action of the
non-perturbative saddle, and the precise pattern of growth
determines a quantity called the Stokes constant. They govern the
non-perturbative imaginary part of the
observable~\cite{Bogomolny:1980ur, ZinnJustin:2004ib, ZinnJustin:2004cg}.
In recent years, the framework of resurgent trans-series has been
developed systematically for quantum mechanics and quantum field
theory~\cite{Dunne:2013ada, Dunne:2014bca, Dorigoni:2014hea,
Aniceto:2018bis}, establishing that the full non-perturbative
content of a theory is, in principle, encoded in its perturbative
data.

Most of these developments have focused on zero-temperature spectral problems
in quantum mechanics, or on quantum field theories on $S^1$ with 
twisted boundary conditions imposed where fractional instantons and bions make 
the resurgence structure accessible semiclassically~\cite{Dunne:2012ae, Dunne:2012zk}. 
Resurgence has also been applied to large-$N$ phase transitions, 
such as the Gross--Witten--Wadia unitary matrix model~\cite{Gross:1980he,Wadia:1980cp}, 
in which the trans-series reorganizes as the 't Hooft coupling crosses
its critical value~\cite{Ahmed:2017lhl}. 
Finite-temperature effects on Borel singularities have been
explored in the context of thermal
renormalons~\cite{Loewe:2000, Cavalcanti:2018}, and a recent
study of the double-well partition function at finite Euclidean
time has demonstrated the importance of exact saddle points
beyond the dilute instanton gas~\cite{Dersy:2026ncf}.
However, a direct resurgence-theoretic account of the thermal
bounce-to-sphaleron transition in metastable quantum mechanics
remains largely unexplored.

In such systems, the Euclidean time circle has size~$\beta$,
and the semiclassical saddle controlling the transition rate
changes with temperature, interpolating between a periodic
bounce at low temperature and a static sphaleron
configuration at high temperature.  This raises a natural
question.  Does the perturbative expansion of the thermal free
energy know about the thermal transition?  Can one read off the transition temperature, and
distinguish the second-order from the first-order case, purely
from the large-order behavior of perturbation theory?

In this paper, we address these questions affirmatively.
We study two quantum-mechanical models, one for each class of
thermal transition.  The first is the cubic potential
$V(y) = \frac12 y^2 - \frac13 y^3$, whose period map is monotonic
and the transition is second order, occurring at
$T_c = (2\pi)^{-1}$.  The second is the quintic
potential $V(y) = \frac12 y^2 - \frac52 y^3 + 5 y^4 - 4 y^5$, whose
period map develops a fold.  Two distinct bounce solutions then
coexist in an intermediate temperature range, and the transition is
first order, occurring at $T_c \simeq 0.0878$.  For each model we
compute the perturbative expansion of the thermal free energy
$F(\beta, \lambda)$ in $\lambda = g^2$ up to order $\lambda^{250}$,
combining the Bender--Wu recursion~\cite{Bender:1969si, Bender:1973rz}
for the Rayleigh--Schr\"odinger series of individual energy levels
with the thermal trace over the lowest $100$ energy levels above the vacuum.  The Borel
data $(A, b, S)$ are then extracted at each temperature from
high-order Borel--Pad\'e approximants.

For the cubic model, the extracted Borel singularity $A(T)$ tracks
the on-shell action of the dominant saddle, the bounce on the cold
side and the sphaleron on the hot side, to within about $0.05\%$ over our temperature window, and it inherits the tangential join of
the two branches at $T_c$.  The characteristic exponent takes the value
$b = 1/2$ in the bounce regime and $b = 0$ in the sphaleron regime,
reflecting the translational zero mode which exists only around the
bounce.  The Stokes constant, extracted from the same perturbative
data, reproduces the one-loop prefactor of
Affleck~\cite{Affleck:1980ac} to better than a percent away from $T_c$.

For the quintic model, the same analysis shows qualitatively 
different behavior.  The Borel singularity $A(T)$ now develops a
kink at $T_c \simeq 0.0878$, where the bounce and sphaleron actions
cross transversally, unlike the smooth join observed
in the cubic potential example. On each branch it still reproduces the dominant on-shell action to within $0.03\%$ across our temperature window. The characteristic exponent $b$ again moves between $1/2$ and
$0$ across the transition, and the Stokes constant follows the
one-loop prefactor of the dominant saddle on each side.  
The first-order nature of the transition thus leaves a clear imprint on the Borel plane
as a kink in $A(T)$.

These results demonstrate that the perturbative expansion of the
thermal free energy around the false vacuum knows the thermal
transition in complete detail. The transition temperature, the
order of the transition, and even the one-loop data, including the number of zero modes, around the
dominant saddle can all be recovered from the large-order
coefficients without relying on semiclassical inputs.  The extraction becomes
delicate only in a narrow window around $T_c$, where the two
leading Borel singularities are nearly degenerate and the
single-saddle asymptotics converge slowly.  
%A systematic treatment
%of this two-saddle interference, presumably requiring uniform
%large-order asymptotics, is left for future work.

Several directions deserve further study.  
It would be desirable to understand our
observations within the exact WKB framework where the Borel
singularities are identified with periods of the WKB differential
on the spectral curve. In particular, clarifying what the fold of the period map and the first-order transition correspond to in this language would put our numerical observations on a rigorous mathematical footing.
Extending the analysis to systems with more degrees of freedom, such as coupled oscillators
and matrix quantum mechanics, would be a natural next step toward
thermal transitions in large-$N$ matrix models and quantum field
theories.

In field theory, and especially in the presence of gravity, the one-loop prefactors of vacuum decay rates are notoriously subtle. For the Coleman–De Luccia and Hawking–Moss instantons \cite{Coleman:1980aw, Hawking:1981fz}, the fluctuation determinants involve delicate issues of negative modes and gauge fixing, and their evaluation has so far been carried out only in limiting cases \cite{Ivo:2025fwe}. Our analysis suggests an alternative approach.  Since the
full one-loop prefactor can be recovered from 
sufficiently many perturbative coefficients around
the false vacuum, a high-order perturbative computation may
determine these determinants without ever solving the fluctuation
problem around the non-perturbative saddle.  Whether such a program
can be carried out in practice, even in a truncated or
minisuperspace setting, is an interesting open problem.

The rest of the paper is organized as follows.
Section~\ref{sec:review} reviews the semiclassical theory of
thermal vacuum decay and the elements of Borel resurgence used in
this work.  Section~\ref{sec:model} introduces the two models,
explains the high-order perturbative computation of the thermal
free energy, and describes how the Borel data $(A, b, S)$ are
extracted.  Section~\ref{sec:results} presents the numerical results
and compares them with the semiclassical predictions.

%======================================================================
\section{Preliminaries}
\label{sec:review}
%======================================================================

\subsection{Thermal vacuum decay and competing saddles}
\label{sec:thermal}

Consider a quantum-mechanical system with a metastable potential
$V(x)$ that has a local minimum at $x=0$ and a barrier of
height $V_{\mathrm{top}}$. 
The thermal free energy can be defined as a Euclidean path
integral on $S^1$ below
\begin{align}
e^{-\beta F(\beta)} \;=\; \int \mathcal{D}x(\tau) \;
\exp\!\Big[-\frac{1}{\hbar}\,S_E\big[x(\tau)\big]\Big]\,,
\label{eq:pathint}
\end{align}
where the Euclidean action is $S_E = \int_0^{\beta\hbar} d\tau ~ \big( \frac12 {\dot x}^2 + V(x) \big)$. 
Here we impose the periodic boundary condition $x(\tau+\beta \hbar) = x(\tau)$.
It was argued in \cite{Affleck:1980ac} 
that the false-vacuum decay rate at temperature $T = 1/\beta$ is encoded 
in the imaginary part of the thermal free energy,
\begin{align}
\Gamma \;=\; -\,\frac{2}{\hbar}\;\mathrm{Im}\,F(\beta)\,.
\label{eq:decayrate}
\end{align}
 
In the semiclassical limit $\hbar \to 0$, 
the path integral is dominated by saddle points of~$S_E$,   
\begin{align}
    - \frac{d^2}{d\tau^2} x(\tau) + V'\big(x(\tau)\big) = 0 \ .
\label{eq:saddleeq}
\end{align}
For a metastable potential, two types of non-trivial saddle compete.
 
The first is the \emph{bounce}, a non-constant periodic solution
of \eqref{eq:saddleeq} with period~$\beta$. Such a solution $\bar x(\tau)$
oscillates back and forth in the inverted potential $(-V)$ at a
classical energy $(-E)$ with $E<V_{\mathrm{top}}$,
\begin{align}
    \frac12 \left( \frac{d}{d\tau} \bar x(\tau)\right)^2 - V \big(\bar x(\tau)\big)  = - E \ .  
\end{align}
One can argue that the period of the motion is identified as $\beta$,
\begin{align}
    \beta = \frac{2}{\hbar} \int_{x_l}^{x_r} dx ~ \frac{1}{\sqrt{2(V(x) - E(\beta) )}}\ ,
\label{eq:periodmap}
\end{align}
where $x_l,x_r$ denote the turning points. The bounce has a finite on-shell 
Euclidean action, 
\begin{align}
S_b(\beta) \;=\;
2 \int_{x_l}^{x_r} dx ~\sqrt{2(V(x) - E(\beta) )}  \;+\; \beta\hbar \,E \, 
\;\equiv\; B(\b) + \beta\hbar \; E(\beta) \, .
\label{eq:bounceaction}
\end{align} 
At low temperature, the bounce
describes quantum tunneling through the barrier at
energy $E$.
 
The second saddle is the \emph{sphaleron}, the static solution
$x(\tau)= x_{\mathrm{top}}$ that sits at the top of the barrier 
$V(x_{\mathrm{top}})=V_{\mathrm{top}}$ for all Euclidean time.
Its action is then simply
\begin{align}
S_{\mathrm{sph}} \;=\; \beta\hbar\,V_{\mathrm{top}}\,.
\label{eq:sphaction}
\end{align}
At high temperature, the sphaleron
dominates and the decay proceeds by classical thermal jump
over the barrier.

Both saddle points possess exactly one negative mode in the
spectrum of fluctuations around them.  This single negative
eigenvalue leads to a factor of~$i$ in the Gaussian
integration around the saddle, which gives rise to the imaginary part
in \eqref{eq:decayrate} and hence of the physical decay rate. 
In short, the (one-loop) decay rates in the two regimes are
\begin{align}
    \Gamma_b = \left( \frac{B(\beta)}{2\pi \hbar} \right)^{1/2}\;
    \left| \frac{\det\big[ - \partial_\t^2 + \omega_0^2 \big]}
    {\det' \big[ - \partial_\t^2 + V''(\bar x(\tau))\big]}
    \right|^{1/2} e^{-S_b(\beta)/\hbar }\ ,
\label{eq:bounce_rate}
\end{align}
for the bounce and
\begin{align}
    \Gamma_{\mathrm{sph}} = \frac{\omega_{\mathrm{top}}}{2\pi}\;
    \left[ \frac{\sinh\!\big(\beta \hbar\omega_0/2\big)}
    {\sin\!\big(\beta\hbar\omega_{\mathrm{top}}/2\big)} \right]
    e^{-S_{\mathrm{sph}}/\hbar}\ ,
\label{eq:sph_rate}
\end{align}
for the sphaleron. Here $\omega_0 = \sqrt{V''(0)}$ is the frequency of the false vacuum and $\omega_{\mathrm{top}} = \sqrt{|V''(x_{\mathrm{top}})|}$ is the
magnitude of the imaginary frequency at the barrier top.

We note that the bounce, unlike the sphaleron, has
a zero mode associated with translation symmetry along the
Euclidean circle. In \eqref{eq:bounce_rate}, $\det'$  denotes the functional
determinant with zero eigenvalue removed. 
The integration over this collective coordinate
gives rise to the factor $(B(\beta)/2\pi\hbar)^{1/2}$ in \eqref{eq:bounce_rate}, 
and contributes an extra power of $\hbar^{-1/2}$ to the one-loop
prefactor.  No such factor appears in the decay rate for sphaleron \eqref{eq:sph_rate}.
As we will see in Section~\ref{sec:resurgence}, this difference
in the $\hbar$-dependence of the prefactor directly determines
the large-order growth of perturbative coefficients.
The characteristic exponent $b$ takes the value $b = 1/2$ in the
bounce regime, reflecting the extra $\hbar^{-1/2}$ from the
zero mode, and $b = 0$ in the sphaleron regime where no such
mode exists.

The transition between the two regimes can be characterized in terms of the
period map $\beta(E)$ in~\eqref{eq:periodmap}.  As $E$ increases from the
bottom of the false vacuum toward the barrier top, $\beta$
starts large and eventually approaches a limiting value
$\beta_\mathrm{top} = 2\pi/(\hbar \omega_{\mathrm{top}})$.

If $\beta(E)$ decreases monotonically as $E$
increases toward~$V_{\mathrm{top}}$, then at every temperature
above~$T_c = T_\mathrm{top} = \omega_{\mathrm{top}}/(2\pi)$ the bounce solution ceases to
exist and the sphaleron takes over smoothly.  The bounce action
$S_b(\beta)$ joins the sphaleron action
$\beta V_{\mathrm{top}}$ continuously at~$\beta_c$.
This is a second-order transition in the language of
Chudnovsky~\cite{Chudnovsky:1992xrj}, where the transition 
temperature $T_c$ is equal to $T_\mathrm{top}$.

If instead $\beta(E)$ develops a fold, meaning it first
decreases and then turns back, two distinct bounce solutions
at different energies coexist for a given temperature in an
intermediate range. Since $dS_b(\beta)/d\beta = E(\beta)$, 
the bounce with lower energy has the
smaller on-shell action and dominates over the one with higher energy.
The dominant saddle jumps discontinuously from the bounce to the sphaleron
at the transition temperature~$T_c$ determined by the equality
$S_b(\beta_c) = \beta_c V_{\mathrm{top}}$.
This is a first-order transition.  The transition
temperature~$T_c$ is generally different from~$T_\mathrm{top}$.

%----------------------------------------------------------------------
\subsection{Borel resurgence}
\label{sec:resurgence}
%----------------------------------------------------------------------
 
Perturbative expansions in quantum mechanics are generically
asymptotic.  The coefficients grow factorially at large order,
and the series has zero radius of convergence.  Optimal
truncation at the smallest term still leaves an error of order
$e^{-A/\lambda}$, where $A$ is a non-perturbative scale set by the
nearest saddle point~\cite{Bender:1973rz}.  Borel resummation
provides a systematic way to go beyond this
limit~\cite{ZinnJustin:2004ib, Dunne:2014bca,
Dorigoni:2014hea, Aniceto:2018bis, Marino:2015yie}.
We review the essential background below. 
 
Let us start with a certain physical observable which admits an asymptotic expansion,
\begin{align}
\Phi(\lambda) \;\sim\; \sum_{n=0}^{\infty} a_n \, \lambda^n 
\qquad (\lambda\to 0)
\label{eq:asymptotic}
\end{align}
whose coefficients grow as
\begin{align}
a_n \;\sim\; \frac{S}{2\pi}\,
\frac{\Gamma(n + b)}{A^{n+b}}
\qquad (n \to \infty) \,.
\label{eq:largeorder}
\end{align}
Here we assume that $A$ is a positive constant, $b$ is a real characteristic exponent, and $S$ is a constant prefactor. The large-order behavior of the form~\eqref{eq:largeorder} 
suffices for the present work. 

The generalized Borel transform of index~$b$ defined by
\begin{align}
\widehat{\mathcal{B}}(t) \;=\;
\sum_{n=0}^{\infty}
\frac{a_n}{\Gamma(n+b)}\, t^n \, 
\label{eq:boreltransform}
\end{align}
removes the factorial growth, which makes the series 
$\widehat{\mathcal{B}}(t)$ convergent in a disk of 
radius~$A$ around the origin. Provided that $\widehat{\mathcal{B}}(t)$ 
has no singularity on the positive real axis, 
one can recover the original function via the Laplace-type integral, 
\begin{align}
\mathcal{S}\Phi(\lambda) \;=\;
\int_0^{\infty} dt\;
e^{-t}\,t^{b-1}\,\widehat{\mathcal{B}}(\lambda t) \,.
\label{eq:borelsum}
\end{align}
This is because the identity
$\int_0^{\infty} e^{-t}\,t^{n+b-1}\,dt
= \Gamma(n+b)$ reconstructs each term $a_n\,\lambda^n$.
This procedure is known as the Borel resummation of~$\Phi(\lambda)$.

We however notice that $\widehat{\mathcal{B}}(t)$ has a
singularity at $t = A$ on the positive real axis. 
To see this, we substitute the large-order
form~\eqref{eq:largeorder} into~\eqref{eq:boreltransform}
to read off the singular part of the Borel transform near $t = A$
\begin{align}
\widehat{\mathcal{B}}^{\,\mathrm{sing}}(t)
\;=\; \frac{S}{2\pi A^b}\,
\sum_{n=0}^{\infty}
\left(\frac{t}{A}\right)^{\!n}
\;=\; \frac{S}{2\pi A^b}\cdot
\frac{1}{1 - t/A} \,.
\label{eq:borelsingularity}
\end{align}
The singularity at $t = A$ is a simple pole on the positive
real axis, and the integral~\eqref{eq:borelsum} becomes ill-defined.

We thus have to provide a prescription to account for the singularity
\eqref{eq:borelsingularity}. 
One natural prescription is to 
deform the contour to pass slightly above or below the pole.  
This gives the lateral Borel resummations $\mathcal{S}_{\pm}$, whose difference
is determined by Cauchy's theorem, 
\begin{align}
\mathcal{S}_+\Phi(\lambda) - \mathcal{S}_-\Phi(\lambda)
\;=\; -i\,S\, \frac{e^{-A/\lambda}}{\lambda^{b}} \,.
\label{eq:ambiguity}
\end{align}
The above ambiguity is closely related to a non-perturbative imaginary contribution
of order $e^{-A/\lambda}\,\lambda^{-b}$.

Since a well-defined physical observable should be free of such ambiguities, 
\eqref{eq:ambiguity} signals that the
perturbative expansion alone is incomplete. This strongly implies the existence 
of the non-perturbative contributions to the observable. A consistent
non-perturbative completion takes the form of a trans-series,
\begin{align}
\Phi(\lambda) \;=\; \sum_{k=0}^{\infty} \sigma^k\,
e^{-kA/\lambda}\, \Phi_k(\lambda) \,,
\label{eq:transseries}
\end{align}
where each $\Phi_k(\lambda) = \sum_n a_n^{(k)}\,\lambda^n$ is itself
an asymptotic series and $\sigma$ is the trans-series
parameter.
The $k=0$ sector is the original perturbative expansion
and $k \geq 1$ sectors describe the non-perturbative 
contributions.
Resurgence is the statement that the Borel transform
of each sector contains information about all other sectors.
In particular, the ambiguity in the Borel resummation
of the perturbative sector $\Phi_0(g)$ has to be canceled by a corresponding ambiguity in the $k=1$ sector, and this cancelation propagates
through all higher sectors, resulting in the unambiguous full
trans-series representation of the physical observable.

The Borel singularity parameters $(A, b, S)$ have direct
physical meaning.  The location~$A$ equals the Euclidean
action of the relevant non-perturbative saddle point.
The characteristic exponent~$b$ encodes the one-loop fluctuation spectrum
around that saddle, including the number of zero modes.
The Stokes constant~$S$ determines the full one-loop
prefactor, incorporating the ratio of fluctuation
determinants between the saddle and the perturbative vacuum.
 
These three quantities can all be extracted numerically
from the perturbative coefficients alone.
Given the asymptotic form~\eqref{eq:largeorder}, the ratio
of successive coefficients grows linearly in~$n$,
\begin{align}
\frac{a_n}{a_{n-1}} \;\sim\; \frac{n+b-1}{A}
\qquad (n \to \infty) \,.
\end{align}
The leading singularity can be therefore extracted from the
sequence
\begin{align}
n\,\frac{a_{n-1}}{a_n}
\;\xrightarrow{n\to\infty}\; A \,.
\label{eq:ratio}
\end{align}
The convergence rate can be accelerated by
Richardson extrapolation, which systematically removes the
$1/n$~corrections.  The characteristic exponent~$b$ and the
Stokes constant~$S$ are then extracted from further
corrections to the ratio.
With perturbative data of sufficiently high order,
all three quantities can be determined to high precision.
 
For the present work, the asymptotic series of our interest is the
perturbative expansion of the thermal free energy
$F(\beta, \lambda)$ around the false vacuum in powers of the
coupling~$\lambda$.  We will show that the Borel transform of this series has a
singularity on the positive real axis, so the free energy
is not Borel summable. The corresponding ambiguity
\eqref{eq:ambiguity} then gives rise to a non-perturbative
imaginary part of the free energy~$F$, which is directly related to the
false-vacuum decay rate \eqref{eq:decayrate}.
The Borel singularity parameters $A$, $b$, and $S$ all
become functions of the inverse temperature~$\beta$.
We will verify that the $\beta$-dependence of these
quantities reveals the thermal transition between the
bounce and the sphaleron.

%======================================================================
\section{The Models and the Perturbative Analysis}
\label{sec:model}
%======================================================================
 
We address the questions raised in Section~\ref{sec:review} with two
specific quantum-mechanical models with metastable potentials.  One is
for the second-order thermal transition between the bounce and the
sphaleron, and the other is for the first-order transition.  For each
model we compute the thermal free energy to high order in perturbation
theory, and investigate how their large-order coefficients behave.
 
%----------------------------------------------------------------------
\subsection{The cubic and quintic models}
\label{sec:models}
%----------------------------------------------------------------------
 
\begin{figure}[t!]
  \centering
  \includegraphics[width=1\textwidth]{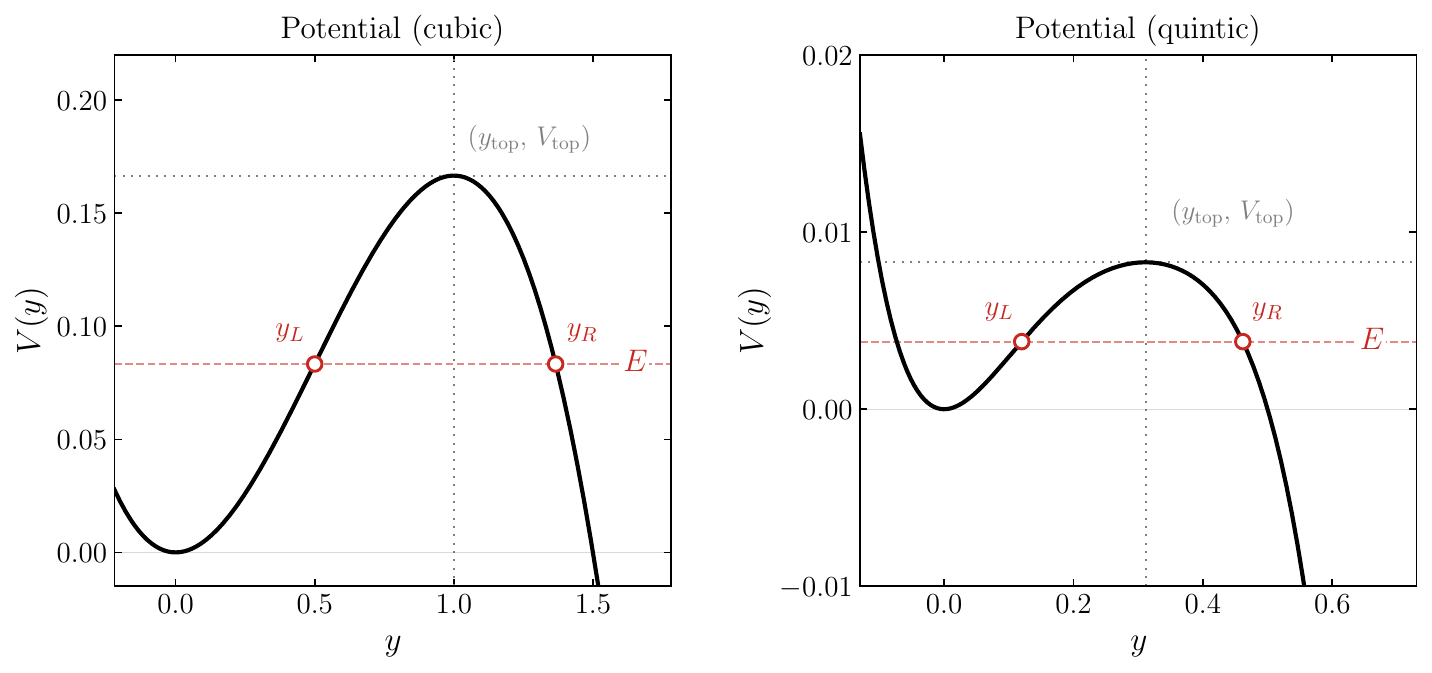}
  \caption{The two metastable potentials.
  \emph{Left}: the cubic potential~\eqref{eq:Vcubic}, with metastable
  minimum at $y=0$ and barrier top at
  $(y_{\mathrm{top}},V_{\mathrm{top}})=(1,1/6)$.  \emph{Right}: the
  quintic potential~\eqref{eq:Vquintic}, with metastable minimum at
  $y=0$ and barrier top at
  $(y_{\mathrm{top}},V_{\mathrm{top}})\simeq(0.311,8.30\times 10^{-3})$.}
  \label{fig:potentials}
\end{figure}

Both of our models describe a particle in a metastable potential at
finite temperature.  With $\hbar=1$, the thermal partition function is
the Euclidean path integral over a circle of circumference~$\hat\beta$,
\begin{align}
S_E \;=\; \int_0^{\hat\beta} d\tau \;
\Big[\, \tfrac12\,\dot x^2 \;+\; \frac{1}{g^2}\,V(g x) \,\Big]\,,
\qquad x(\tau+\hat\beta)=x(\tau)\,,
\label{eq:SEcoupling}
\end{align}
where $g$ is a coupling constant.  For the present work we consider two
different potentials~$V$,
\begin{align}
V_{\mathrm{cubic}}(y) \;&=\; \tfrac12\,y^2 \;-\; \tfrac13\,y^3\,,
\label{eq:Vcubic}\\[2pt]
V_{\mathrm{quintic}}(y) \;&=\;
\tfrac12\,y^2 \;-\; \tfrac52\,y^3 \;+\; 5\,y^4 \;-\; 4\,y^5\,.
\label{eq:Vquintic}
\end{align}
Both are normalized so that the false vacuum sits at the origin $y=0$,
and have a single barrier of finite height~$V_{\mathrm{top}}$.
The shape of each potential is depicted in Figure~\ref{fig:potentials}.

For these potentials the weak-coupling expansion can be understood as
the semiclassical expansion.  To see this, rescale $y = g x$ in
\eqref{eq:SEcoupling}, which turns the coupling into an overall
prefactor of the action,
\begin{align}
S_E \;=\; \frac{1}{g^2}\int_0^{\hat\beta} d\tau \;
\Big[\, \tfrac12\,\dot y^2 \;+\; V(y) \,\Big]
\;\equiv\; \frac{1}{g^2}\,S_0[y]\,.
\label{eq:Srescaled}
\end{align}
The path-integral weight is then given by $e^{-S_0[y]/g^2}$.  Thus $g^2$
plays the role of the effective $\hbar$ of Section~\ref{sec:review} for
the action $S_0[y]$ at inverse temperature $\beta=\hat\beta/g^2$.

Upon the identification $\hbar_\text{eff}\simeq g^2$, the semiclassical results of
Section~\ref{sec:review} can be expressed as follows.  In the bounce
regime,
\begin{align}
2 \left| \mathrm{Im}\,F_{\mathrm{bounce}}(\hat\beta,g^2) \right|
\;=\; \left(\frac{B(\hat\beta)}{2\pi g^2}\right)^{1/2}
\left| \frac{\det\big[-\partial_\tau^2+\omega_0^2\big]}
{\det'\big[-\partial_\tau^2+V''(\bar y)\big]} \right|^{1/2}
e^{-S_b(\hat\beta)/g^2}\,,
\label{eq:ImFbounce}
\end{align}
with
\begin{align}
\begin{split}
\hat\beta  & = 2\int_{y_l}^{y_r}
\frac{dy}{\sqrt{2\,(V(y)-E(\hat \beta))}}\,,
\\
S_b(\hat\beta)  & = 
2\int_{y_l}^{y_r} dy\,\sqrt{2\,(V(y)-E(\hat \beta))} \;+\; \hat\beta\,E(\hat \beta)
= B(\hat \beta) + \hat \beta E(\hat \beta)\,,
\end{split}
\label{eq:periodaction}
\end{align}
where $y_l,y_r$ are the turning points as shown in Figure~\ref{fig:potentials}, and $S_b(\hat\beta)$ is the
full on-shell bounce action.  In the sphaleron regime,
\begin{align}
2 \left| \mathrm{Im}\,F_{\mathrm{sph}}(\hat\beta,g^2)\right|
\;=\; \frac{\omega_{\mathrm{top}}}{2\pi}\,
\left[\frac{\sinh(\hat\beta\,\omega_0/2)}{\sin(\hat\beta\,\omega_{\mathrm{top}}/2)}\right]
e^{-\hat\beta\,V(y_{\mathrm{top}})/g^2}\,,
\label{eq:ImFsph}
\end{align}
with $\omega_0=\sqrt{V''(y=0)}$ and
$\omega_{\mathrm{top}}=\sqrt{|V''(y_{\mathrm{top}})|}$. 
In what follows, we drop the hat and simply write $\beta$  for 
the inverse temperature unless it causes confusion.

\begin{figure}[t!]
  \centering
  \includegraphics[width=1\textwidth]{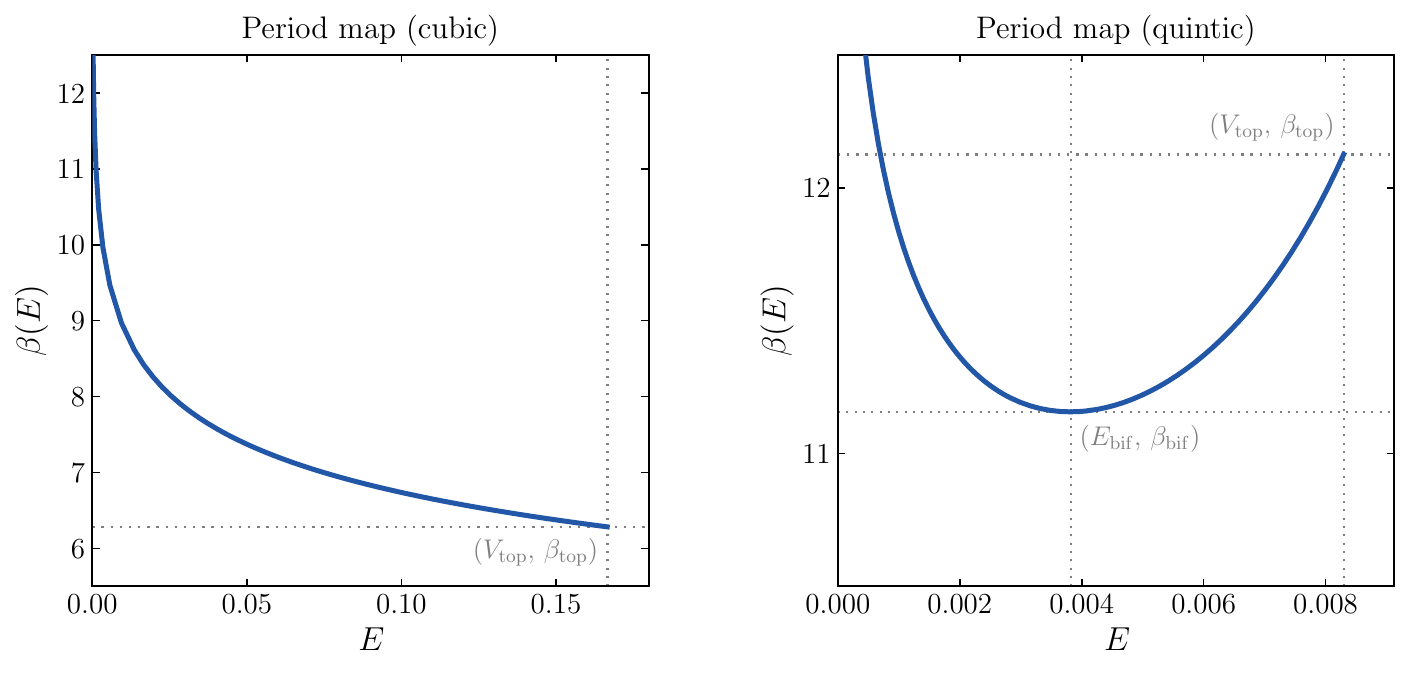}
  \caption{Period maps $\beta(E)$ of the two models.  \emph{Left}
  (cubic): $\beta(E)$ decreases monotonically from infinity at the false
  vacuum to the harmonic limit
  $(V_{\mathrm{top}},\beta_{\mathrm{top}})=(1/6,2\pi)$.  \emph{Right} (quintic):
  $\beta(E)$ is non-monotone, descending past the harmonic limit
  $(V_{\mathrm{top}},\beta_{\mathrm{top}})\simeq(8.30\times 10^{-3},12.13)$
  to the fold bifurcation
  $(E_{\rm bif},\beta_{\rm bif})\simeq(3.82\times 10^{-3},11.16)$
  and rising back to $\beta_{\mathrm{top}}$.}
  \label{fig:periodmaps}
\end{figure}
The cubic potential~\eqref{eq:Vcubic} is a canonical example of the
second-order thermal transition between the bounce and the sphaleron,
while the quintic potential~\eqref{eq:Vquintic} is a simple 
example of the first-order transition. To see this, let us consider 
the period map $\beta(E)$ in~\eqref{eq:periodaction} depicted in Figure~\ref{fig:periodmaps}.  For
the cubic potential $\beta(E)$ decreases monotonically from infinity at
the false vacuum to $\beta_{\mathrm{top}}$ at the barrier top, so the
bounce ceases to exist beyond $\beta_{\mathrm{top}}$ and the transition
occurs exactly at $\beta_c=\beta_{\mathrm{top}}$. Furthermore, 
the bounce and sphaleron branches of the dominant semiclassical action
join tangentially. On the other hand, 
$\beta(E)$ is non-monotonic for the quintic potential. More precisely, 
$\beta(E)$ first descends below $\beta_{\mathrm{top}}$ to a minimum
$\beta_{\rm bif}\simeq 11.16$ at the bifurcation point and turns back up to
$\beta_{\mathrm{top}}$. As a consequence, two distinct bounces coexist for
$\beta\in(\beta_{\rm bif},\beta_{\mathrm{top}})$.  The transition is
then controlled by the condition
$S_b(\beta_c)=\beta_c V_{\mathrm{top}}$, met at $\beta_c\simeq 11.39$
($T_c\simeq 0.0878$), strictly between $\beta_{\rm bif}$ and
$\beta_{\mathrm{top}}$. In this case one can argue that 
bounce and sphaleron branches now cross transversally rather than tangentially.

For later convenience, we present the relevant parameters of the two
models.  Both have $\omega_0=\sqrt{V''(0)}=1$ at the false vacuum.  At
the barrier top, the cubic has $\omega_{\mathrm{top}}=1$,
$V_{\mathrm{top}}=1/6$, and $\beta_{\mathrm{top}}=2\pi$, with the
transition at $\beta_c=\beta_{\mathrm{top}}=2\pi$ ($T_c\simeq 0.159$) while the quintic has
$\omega_{\mathrm{top}}\simeq 0.518$,
$V_{\mathrm{top}}\simeq 8.30\times 10^{-3}$, and
$\beta_{\mathrm{top}}\simeq 12.13$, with the transition at
$\beta_c\simeq 11.39$ ($T_c\simeq 0.0878$).

Equations~\eqref{eq:ImFbounce}--\eqref{eq:ImFsph} tell us what to
expect from the perturbative expansion of $F(\beta,g^2)$, if its
large-order behavior indeed encodes the semiclassical physics.  Read
through the Borel dictionary~\eqref{eq:largeorder}, we expect the
leading Borel singularity to sit at the action of the dominant saddle,
\begin{align}
A(\beta) \;=\; \min\big\{\, S_b(\beta),\;
\beta V_{\mathrm{top}} \,\big\}\,,
\label{eq:leastaction}
\end{align}
the characteristic exponent to take the value $b=\tfrac12$ in the bounce
regime and $b=0$ in the sphaleron regime, and the Stokes
constant~$S(\beta)$ to reproduce the one-loop prefactor of
the corresponding saddle~\eqref{eq:ImFbounce}--\eqref{eq:ImFsph},
\begin{align}
S_{\mathrm{th}}(\beta)/2\pi
=
\begin{cases}
\displaystyle
\left(\frac{B(\beta)}{2\pi}\right)^{1/2}
\left|
\frac{\det\!\left[-\partial_\tau^2+\omega_0^2\right]}
{\det'\!\left[-\partial_\tau^2+V''(\bar y)\right]}
\right|^{1/2}
& (\text{bounce regime})\,,
\\[14pt]
\displaystyle
\frac{\omega_{\mathrm{top}}}{2\pi}
\left[
\frac{\sinh(\beta\omega_0/2)}
{\sin(\beta\omega_{\mathrm{top}}/2)}
\right]
& (\text{sphaleron regime})\,.
\end{cases}
\label{eq:Saff}
\end{align}
The remainder of the paper examines these three predictions, temperature by
temperature, against the large-order behavior of perturbation theory.

%----------------------------------------------------------------------
\subsection{High-order perturbative free energy}
\label{sec:pertF}
%----------------------------------------------------------------------

We now explain how the perturbative coefficients of the thermal free
energy are computed.  We use
$\lambda=g^2$ as the expansion parameter and write
\begin{align}
 Z(\beta,\lambda)=\sum_{\nu\geq 0} e^{-\beta E_\nu(\lambda)}\, ,   
\end{align}
and 
\begin{align}
F(\beta,\lambda)
\equiv -\frac{1}{\beta}\log Z(\beta,\lambda)
=\sum_{n\geq 0} F_n(\beta)\lambda^n\, .
\label{eq:Fseries}
\end{align}
Here $F_0(\beta)$ is the free energy of the harmonic oscillator.  The
coefficients $F_n(\beta)$ with $n\geq 1$ are the perturbative data
from which the Borel singularity data
$(A,b,S)$ will be read off.

The computation proceeds in two steps.  First, for each oscillator level
\(\nu\), we compute the Rayleigh--Schrödinger perturbative expansion
\begin{align}
E_\nu(\lambda)
=
\left(\nu+\frac12\right)
+\sum_{n\geq 1}\varepsilon_{\nu,n}\lambda^n .
\label{eq:Enseries}
\end{align}
The coefficients $\varepsilon_{\nu,n}$ are obtained by the
Bender--Wu recursion \cite{Bender:1969si,Sulejmanpasic:2016fwr}, implemented in arbitrary-precision arithmetic.
This method is particularly useful for the present problem since
it produces very high orders of perturbation theory around a locally
harmonic oscillator vacuum.

Second, we assemble the thermal trace as a formal power series in
$\lambda=g^2$.  For each level we expand
\begin{align}
e^{-\beta E_\nu(\lambda)}
=
e^{-\beta(\nu+1/2)}
\exp\!\left[-\beta\sum_{n\geq 1}
\varepsilon_{\nu,n}\lambda^n\right]
\end{align}
to the desired order, sum over $\nu$, and finally take the logarithm
as a formal series.  This gives $F_n(\beta)$ directly from the table
of energy corrections $\varepsilon_{\nu,n}$.

In numerical practice, the calculation is controlled by two cutoffs.  The first
is the perturbative order $n_{\max}$, and the second is the number of
levels $\nu_{\max}$ involved in the thermal trace.  We compute
$E_\nu(\lambda)$ for $0\leq \nu\leq \nu_{\max}$ and
$0\leq n\leq n_{\max}$, and then construct
$F_n(\beta)$ for the same range of $n$.  In the production runs we
take
$$
n_{\max}=250,\qquad
\nu_{\max}=100
$$
for both the cubic and quintic models.

These cutoffs are carefully chosen for the large-order analysis, since the high-order
coefficients $F_n(\beta)$ receive non-negligible contributions from
excited levels even when their Boltzmann weights are small.  We
therefore choose $\nu_{\max}$ large enough so that the extracted
large-order data remain stable under increasing the number of levels.

%----------------------------------------------------------------------
\subsection{Extracting the Borel data \(A\), \(b\), and \(S\)}
\label{sec:extraction}
%----------------------------------------------------------------------

We now describe how the Borel data can be extracted from the high-order
thermal coefficients.  We expect that the large-order coefficients behave as 
\begin{align}
F_n(\beta)
\sim
- \frac{S(\beta)}{2\pi}
\frac{\Gamma(n+b(\beta))}{A(\beta)^{\,n+b(\beta)}}
\qquad (n\to\infty).
\label{eq:Fnlargeorder}
\end{align}
The position \(A(\beta)\) of the leading Borel singularity, the
characteristic exponent \(b(\beta)\), and the Stokes constant \(S(\beta)\) can all be read
off from the large-\(n\) behavior of \(F_n(\beta)\).
We observe that the coefficients $F_n(\beta)$
have a fixed sign at large order. This is consistent with the fact that a leading
Borel singularity is located on the positive real axis, responsible for the imaginary part
of the thermal free energy.

The Borel--Pad\'e method provides an efficient way to extract the leading Borel singularity 
$A(\beta)$ from the high-order coefficients $F_n(\beta)$. 
We first perform the analytic continuation of \(\mathcal{B}(t)=\sum_n F_n(\beta)\,t^n/n!\)
by a diagonal Pad\'e approximant \([m/m]\) in \(t\)~\cite{Graffi:1970erh,Baker:1996,Caliceti:2007ra}. 
The location of $A(\beta)$ is then identified as the smallest 
stable positive real pole of the Pad\'e sequence.
%in most of the temperature range considered below. 
To obtain the characteristic exponent $b(\beta)$, we apply the standard 
Dlog--Pad\'e construction~\cite{Guttmann:1989,Baker:1996} to 
the logarithmic derivative of $\mathcal{B}(t)$, 
\begin{align}
b(\beta)=-\operatorname*{Res}_{t=A(\beta)}\frac{d}{dt}\log{\mathcal{B}}(t),
\label{eq:bdlog}
\end{align}
where the residue is evaluated at the theoretical 
value of the corresponding dominant saddle $A(\beta)=A_{\rm th}(\beta)$. The Stokes constant in turn follows directly from 
the generalized Borel transform $\widehat{\mathcal{B}}(t)$~\eqref{eq:boreltransform},  
\begin{align}
S(\beta)=2\pi\,A(\beta)^{\,b-1}\operatorname*{Res}_{t=A(\beta)}
\widehat{\mathcal{B}}(t).
\label{eq:Sres}
\end{align}
Here $(A,b)$ are fixed by their theoretical values 
$(A_{\rm th},b_{\rm th})$ of the corresponding saddle. 
In practice, the diagonal Pad\'e approximant  \([m/m]\)  of 
\(\widehat{\mathcal{B}}(t)\) is used to 
obtain the Stokes constant. 
We compare the numerical value of $S(\beta)$ with the one-loop prefactor $S_{\rm th}(\beta)$ \eqref{eq:Saff}.  
The useful quantity is therefore 
\begin{align}
\frac{S(\beta)}{S_{\rm th}(\beta)} ,
\label{eq:Sratio}
\end{align}
which should approach unity whenever a single saddle dominates the large-order behavior.
Note that each of the Borel data \((A,b,S)\) is taken as the median over the window of diagonal orders
\(m=100,\dots,125\) up to the \([125/125]\) approximant.

%======================================================================
\section{Numerical Results}
\label{sec:results}
%======================================================================

% We now confront the Borel data $(A,b,S)$ extracted from each model's perturbative free energy with the semiclassical expectations on a uniform temperature grid. 

\subsection{Cubic potential}

\begin{figure}[t!]
\centering
\includegraphics[width=0.8\textwidth]{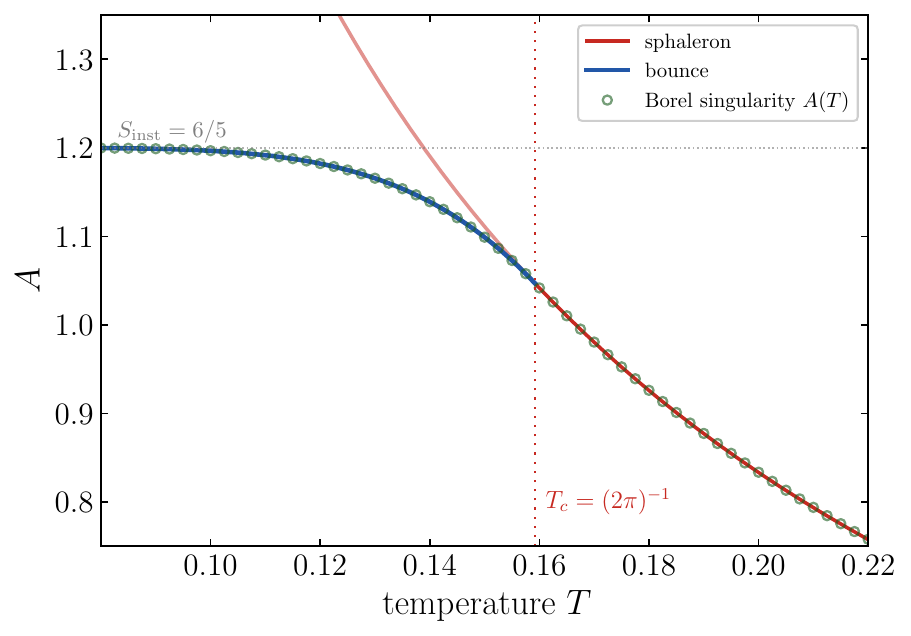}
\caption{Cubic potential. Borel singularity $A(T)$ compared with the classical sphaleron action $\beta V_{\rm top}$ and bounce action $S_b(T)$. The two branches join tangentially at $T_{c}=T_{\rm top}=(2\pi)^{-1}\approx 0.159$.}
\label{fig:cubicAm}
\end{figure}
For the model with the cubic potential~\eqref{eq:Vcubic}, we plot in Figure~\ref{fig:cubicAm} the on-shell actions of the sphaleron and bounce saddles. In the semiclassical limit, 
the saddle with the smaller on-shell action gives the dominant contribution to 
the decay rate of the false vacuum.  The bounce action $S_b(T)$~\eqref{eq:bounceaction}
controls the low-temperature regime and approaches the zero-temperature 
instanton value $6/5$ as  $T\to 0$. On the other hand, the sphaleron action 
$\beta V_{\rm top}=V_{\rm top}/T$~\eqref{eq:sphaction} controls the high-temperature regime and scales as $1/T$. 
The two branches meet at  $T_{c}=T_{\rm top}=(2\pi)^{-1}\approx 0.159$, where they share a common tangent. Thus, as discussed in Section~\ref{sec:model}, the cubic potential~\eqref{eq:Vcubic} exhibits the second-order transition.

The open circles in Figure~\ref{fig:cubicAm} show the Borel singularity $A(T)$ obtained solely from the perturbative coefficients computed up to a large order. They lie on the sphaleron line $\beta V_{\mathrm{top}}$ on the hot side $T>T_c$, on the periodic-bounce line $S_b(T)$ on the cold side $T<T_c$, and approach the instanton value $6/5$ at the coldest grid point. The extracted $A(T)$ joins smoothly at $T_c$. 
%We also extract the same quantity $A(T)$ from the poles 
%of the Pad\'e approximant to the generalized Borel transform, shown in the right panel of 
%Figure~\ref{fig:cubicAm}. The resulting data again follow the classical action on both sides of $T_c$, including the tangential join. 

Table~\ref{tab:fvst-cubic} compares the extracted $A$ with the corresponding classical action $A_{\rm th}$.
Their relative gap $\Delta$ is of order $10^{-2}\%$ across the window, rising to at most $\Delta\simeq 0.05\%$. We emphasize that this agreement
uses no semiclassical input, and the Borel--Pad\'e extraction indeed reconstructs the on-shell action
of the corresponding saddle at each temperature from the free-energy coefficients up to large order.
\begin{table}[t]
\centering
\scriptsize
\setlength{\tabcolsep}{3pt}
\renewcommand{\arraystretch}{0.85}
\begin{NiceTabular}[baseline=t,colortbl-like]{cccccc}
\toprule
$T$ & $A_{\rm th}$ & $A$ & $\Delta$ & $b$ & $S/S_{\rm th}$ \\
\midrule
\rowcolor{bnc!12} 0.0800 & 1.199731 & 1.199791 & 0.004976 & 0.500 & 1.000 \\
\rowcolor{bnc!12} 0.0825 & 1.199608 & 1.199668 & 0.004996 & 0.500 & 1.000 \\
\rowcolor{bnc!12} 0.0850 & 1.199439 & 1.199499 & 0.005010 & 0.500 & 1.000 \\
\rowcolor{bnc!12} 0.0875 & 1.199215 & 1.199275 & 0.005029 & 0.500 & 1.000 \\
\rowcolor{bnc!12} 0.0900 & 1.198921 & 1.198983 & 0.005168 & 0.500 & 1.000 \\
\rowcolor{bnc!12} 0.0925 & 1.198542 & 1.198602 & 0.005015 & 0.500 & 1.000 \\
\rowcolor{bnc!12} 0.0950 & 1.198060 & 1.198121 & 0.005144 & 0.500 & 1.000 \\
\rowcolor{bnc!12} 0.0975 & 1.197455 & 1.197515 & 0.004970 & 0.500 & 1.000 \\
\rowcolor{bnc!12} 0.1000 & 1.196707 & 1.196765 & 0.004859 & 0.500 & 1.000 \\
\rowcolor{bnc!12} 0.1025 & 1.195791 & 1.195849 & 0.004899 & 0.500 & 1.000 \\
\rowcolor{bnc!12} 0.1050 & 1.194680 & 1.194738 & 0.004868 & 0.500 & 1.000 \\
\rowcolor{bnc!12} 0.1075 & 1.193347 & 1.193409 & 0.005169 & 0.500 & 1.000 \\
\rowcolor{bnc!12} 0.1100 & 1.191761 & 1.191819 & 0.004800 & 0.500 & 1.000 \\
\rowcolor{bnc!12} 0.1125 & 1.189890 & 1.189953 & 0.005288 & 0.500 & 1.000 \\
\rowcolor{bnc!12} 0.1150 & 1.187698 & 1.187757 & 0.004946 & 0.500 & 1.000 \\
\rowcolor{bnc!12} 0.1175 & 1.185149 & 1.185206 & 0.004747 & 0.500 & 1.000 \\
\rowcolor{bnc!12} 0.1200 & 1.182204 & 1.182264 & 0.005035 & 0.500 & 1.000 \\
\rowcolor{bnc!12} 0.1225 & 1.178823 & 1.178880 & 0.004904 & 0.500 & 1.000 \\
\rowcolor{bnc!12} 0.1250 & 1.174961 & 1.175017 & 0.004842 & 0.500 & 1.000 \\
\rowcolor{bnc!12} 0.1275 & 1.170573 & 1.170633 & 0.005138 & 0.500 & 1.000 \\
\rowcolor{bnc!12} 0.1300 & 1.165614 & 1.165679 & 0.005565 & 0.500 & 1.000 \\
\rowcolor{bnc!12} 0.1325 & 1.160033 & 1.160099 & 0.005705 & 0.500 & 1.000 \\
\rowcolor{bnc!12} 0.1350 & 1.153779 & 1.153846 & 0.005785 & 0.500 & 1.000 \\
\rowcolor{bnc!12} 0.1375 & 1.146800 & 1.146869 & 0.006018 & 0.500 & 1.000 \\
\rowcolor{bnc!12} 0.1400 & 1.139039 & 1.139112 & 0.006408 & 0.500 & 1.000 \\
\rowcolor{bnc!12} 0.1425 & 1.130441 & 1.130514 & 0.006481 & 0.500 & 1.000 \\
\rowcolor{bnc!12} 0.1450 & 1.120945 & 1.121024 & 0.006990 & 0.500 & 1.000 \\
\rowcolor{bnc!12} 0.1475 & 1.110491 & 1.110570 & 0.007060 & 0.500 & 1.000 \\
\rowcolor{bnc!12} 0.1500 & 1.099016 & 1.099094 & 0.007137 & 0.499 & 1.000 \\
\rowcolor{bnc!12} 0.1525 & 1.086453 & 1.086536 & 0.007615 & 0.498 & 1.000 \\
\rowcolor{bnc!12} 0.1550 & 1.072736 & 1.072817 & 0.007500 & 0.493 & 1.000 \\
\rowcolor{bnc!12} 0.1575 & 1.057796 & 1.057873 & 0.007261 & 0.396 & 0.978 \\
\bottomrule
\end{NiceTabular}\hspace{1em}
\begin{NiceTabular}[baseline=t,colortbl-like]{cccccc}
\toprule
$T$ & $A_{\rm th}$ & $A$ & $\Delta$ & $b$ & $S/S_{\rm th}$ \\
\midrule
\rowcolor{sph!12} 0.1600 & 1.041667 & 1.041807 & 0.013514 & 0.258 & 0.609 \\
\rowcolor{sph!12} 0.1625 & 1.025641 & 1.025825 & 0.017957 & 0.066 & 0.921 \\
\rowcolor{sph!12} 0.1650 & 1.010101 & 1.010301 & 0.019810 & 0.030 & 0.967 \\
\rowcolor{sph!12} 0.1675 & 0.995025 & 0.995241 & 0.021765 & 0.017 & 0.982 \\
\rowcolor{sph!12} 0.1700 & 0.980392 & 0.980606 & 0.021840 & 0.012 & 0.988 \\
\rowcolor{sph!12} 0.1725 & 0.966184 & 0.966408 & 0.023200 & 0.008 & 0.992 \\
\rowcolor{sph!12} 0.1750 & 0.952381 & 0.952614 & 0.024518 & 0.006 & 0.994 \\
\rowcolor{sph!12} 0.1775 & 0.938967 & 0.939207 & 0.025585 & 0.005 & 0.995 \\
\rowcolor{sph!12} 0.1800 & 0.925926 & 0.926155 & 0.024782 & 0.004 & 0.996 \\
\rowcolor{sph!12} 0.1825 & 0.913242 & 0.913479 & 0.025936 & 0.004 & 0.997 \\
\rowcolor{sph!12} 0.1850 & 0.900901 & 0.901144 & 0.027000 & 0.003 & 0.997 \\
\rowcolor{sph!12} 0.1875 & 0.888889 & 0.889144 & 0.028746 & 0.003 & 0.997 \\
\rowcolor{sph!12} 0.1900 & 0.877193 & 0.877455 & 0.029902 & 0.002 & 0.998 \\
\rowcolor{sph!12} 0.1925 & 0.865801 & 0.866067 & 0.030747 & 0.002 & 0.998 \\
\rowcolor{sph!12} 0.1950 & 0.854701 & 0.854972 & 0.031772 & 0.002 & 0.998 \\
\rowcolor{sph!12} 0.1975 & 0.843882 & 0.844167 & 0.033750 & 0.002 & 0.998 \\
\rowcolor{sph!12} 0.2000 & 0.833333 & 0.833624 & 0.034931 & 0.002 & 0.998 \\
\rowcolor{sph!12} 0.2025 & 0.823045 & 0.823335 & 0.035154 & 0.002 & 0.998 \\
\rowcolor{sph!12} 0.2050 & 0.813008 & 0.813309 & 0.036982 & 0.001 & 0.998 \\
\rowcolor{sph!12} 0.2075 & 0.803213 & 0.803520 & 0.038181 & 0.001 & 0.999 \\
\rowcolor{sph!12} 0.2100 & 0.793651 & 0.793959 & 0.038778 & 0.001 & 0.999 \\
\rowcolor{sph!12} 0.2125 & 0.784314 & 0.784614 & 0.038307 & 0.001 & 0.999 \\
\rowcolor{sph!12} 0.2150 & 0.775194 & 0.775518 & 0.041803 & 0.001 & 0.999 \\
\rowcolor{sph!12} 0.2175 & 0.766284 & 0.766622 & 0.044107 & 0.001 & 0.999 \\
\rowcolor{sph!12} 0.2200 & 0.757576 & 0.757924 & 0.045982 & 0.001 & 0.999 \\
\rowcolor{sph!12} 0.2225 & 0.749064 & 0.749424 & 0.048085 & 0.001 & 0.999 \\
\rowcolor{sph!12} 0.2250 & 0.740741 & 0.741097 & 0.048143 & 0.001 & 0.999 \\
\rowcolor{sph!12} 0.2275 & 0.732601 & 0.732978 & 0.051443 & 0.001 & 0.999 \\
\rowcolor{sph!12} 0.2300 & 0.724638 & 0.725013 & 0.051774 & 0.001 & 0.999 \\
\bottomrule
\end{NiceTabular}
\caption{Cubic potential. The data underlying Figures~\ref{fig:cubicAm} and~\ref{fig:cubicbS} (transition at $T_c=(2\pi)^{-1}\approx 0.159$). 
%(summed over $100$ oscillator levels) at large-order cutoff $n_{\max}=250$. 
$A$, $b$, and $S/S_{\rm th}$ are the Borel--Pad\'e values and $\Delta=|A-A_{\rm th}|/A_{\rm th}$ (\%). 
%Rows are shaded by phase: \textcolor{bnc}{bounce} ($T<T_c$) and \textcolor{sph}{sphaleron} ($T>T_c$).
}
\label{tab:fvst-cubic}
\end{table}
\begin{figure}[t!]
\centering
\begin{minipage}[c]{0.5\textwidth}
  \centering
  \includegraphics[width=\textwidth]{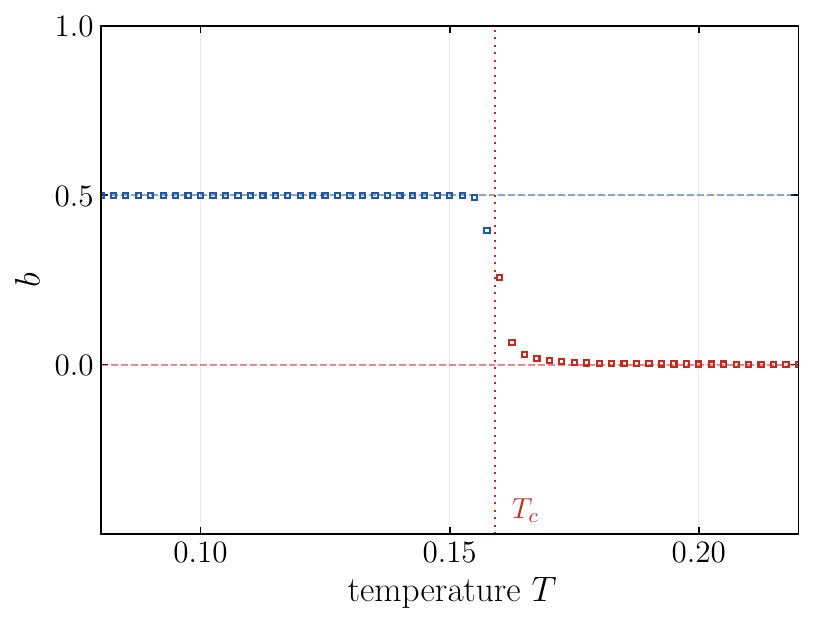}
\end{minipage}\hfill
\begin{minipage}[c]{0.5\textwidth}
  \centering
  \includegraphics[width=\textwidth]{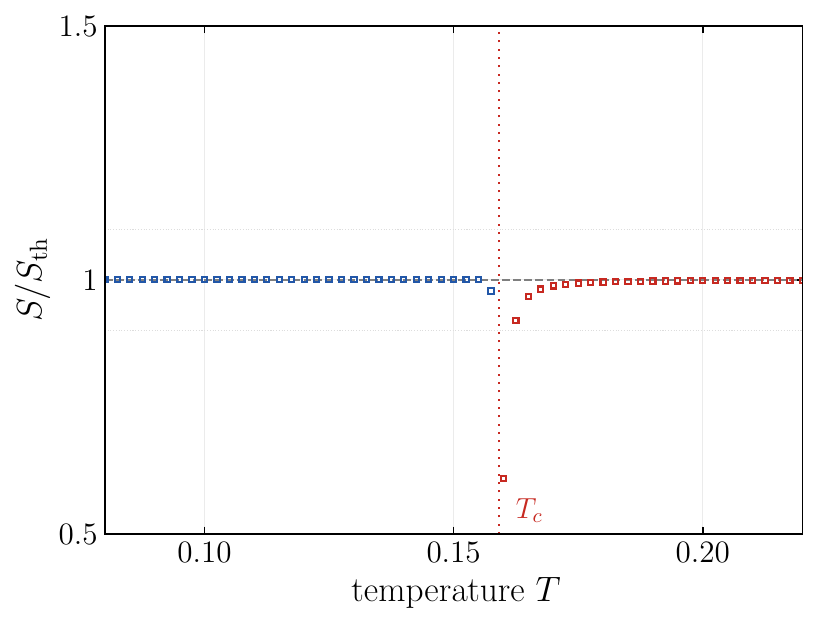}
\end{minipage}
\caption{Cubic potential. \emph{Left:} characteristic exponent $b(T)$ against the theoretical values $0$ and $1/2$ (dashed). \emph{Right:} Stokes ratio $S/S_{\rm th}$, near unity away from $T_{c}$ on both sides and dipping near $T_{c}$.}
\label{fig:cubicbS}
\end{figure}
The left panel of Figure~\ref{fig:cubicbS} shows that the characteristic exponent $b(T)$ 
sits at $b=0$ on the hot sphaleron side while at $b=1/2$ on the cold
bounce side with only a mild excursion near $T_c$ where the single-saddle asymptotics
converge most slowly. These estimates agree with the values expected from the number of zero modes around the corresponding saddle. Thus, $b(T)$ independently confirms the nature of the saddle identified by $A(T)$, and deviates from these values only near $T_{c}$ where the two saddle contributions become comparable. 
  
The right panel of Figure~\ref{fig:cubicbS} shows the Stokes ratio $S/S_{\rm th}$~\eqref{eq:Sratio}. Away from $T_{c}$ it remains at unity to within a few percent on both sides, so the numerical Stokes constant reproduces the Affleck prefactor in~\eqref{eq:Saff}. The ratio $S/S_{\rm th}$ shows deviations from unity only near $T_{c}$ as we expect. Table~\ref{tab:fvst-cubic} lists $b$ and $S/S_{\rm th}$ at each grid point.

\subsection{Quintic potential}

\begin{figure}[tp!]
\centering
\includegraphics[width=0.8\textwidth]{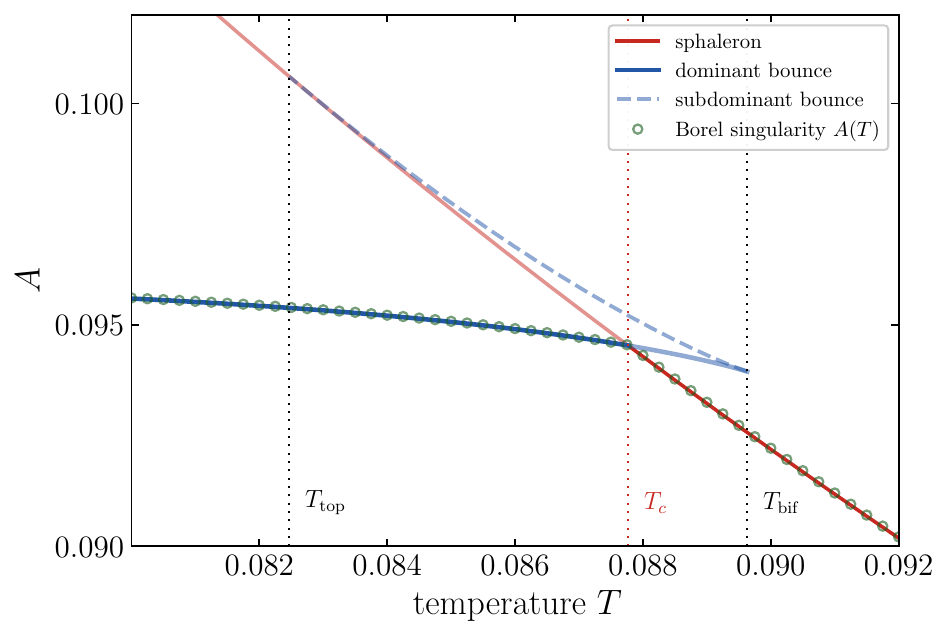}
\caption{Quintic potential. Borel singularity $A(T)$, compared with the classical sphaleron action $\beta V_{\rm top}$ and the dominant bounce action $S_b(T)$, which cross transversally at $T_{c}\simeq 0.0878$. The dashed branch is the subdominant bounce, present between $T_{\rm top}$ and $T_{\rm bif}$.}
\label{fig:quinticAm}
\end{figure}

For the model with the quintic potential \eqref{eq:Vquintic}, Figure~\ref{fig:quinticAm} shows two branches of bounce saddles, originating from the period map $\beta(E)$ with a fold in  Figure~\ref{fig:periodmaps}. Between $T_{\rm top}\simeq0.0825$ and $T_{\rm bif}\simeq0.0896$, two bounce solutions coexist. The dominant bounce branch is described as a solid line while the subdominant branch as a dashed line. In the high temperature regime, the sphaleron action $\beta V_{\rm top}$ \eqref{eq:sphaction} dominates. The sphaleron and the dominant branch of bounce actions cross transversally at $T_c\simeq0.0878$, producing a kink rather than a tangential join. As discussed in Section~\ref{sec:model}, the quintic potential therefore exhibits a first-order transition.

\begin{table}[tp!]
\centering
\scriptsize
\setlength{\tabcolsep}{3pt}
\renewcommand{\arraystretch}{0.85}
\begin{NiceTabular}[baseline=t,colortbl-like]{cccccc}
\toprule
$T$ & $A_{\rm th}$ & $A$ & $\Delta$ & $b$ & $S/S_{\rm th}$ \\
\midrule
\rowcolor{bnc!12} 0.07500 & 0.095845 & 0.095858 & 0.01311 & +0.500 & 0.999 \\
\rowcolor{bnc!12} 0.07525 & 0.095837 & 0.095849 & 0.01284 & +0.500 & 0.999 \\
\rowcolor{bnc!12} 0.07550 & 0.095828 & 0.095839 & 0.01207 & +0.500 & 0.999 \\
\rowcolor{bnc!12} 0.07575 & 0.095819 & 0.095831 & 0.01286 & +0.500 & 0.999 \\
\rowcolor{bnc!12} 0.07600 & 0.095809 & 0.095823 & 0.01461 & +0.500 & 0.999 \\
\rowcolor{bnc!12} 0.07625 & 0.095800 & 0.095812 & 0.01309 & +0.500 & 0.999 \\
\rowcolor{bnc!12} 0.07650 & 0.095789 & 0.095801 & 0.01215 & +0.500 & 0.999 \\
\rowcolor{bnc!12} 0.07675 & 0.095779 & 0.095793 & 0.01467 & +0.500 & 0.999 \\
\rowcolor{bnc!12} 0.07700 & 0.095768 & 0.095781 & 0.01341 & +0.500 & 0.999 \\
\rowcolor{bnc!12} 0.07725 & 0.095756 & 0.095768 & 0.01236 & +0.500 & 0.999 \\
\rowcolor{bnc!12} 0.07750 & 0.095744 & 0.095757 & 0.01325 & +0.500 & 0.999 \\
\rowcolor{bnc!12} 0.07775 & 0.095732 & 0.095745 & 0.01371 & +0.500 & 0.999 \\
\rowcolor{bnc!12} 0.07800 & 0.095719 & 0.095732 & 0.01316 & +0.500 & 0.999 \\
\rowcolor{bnc!12} 0.07825 & 0.095706 & 0.095719 & 0.01389 & +0.500 & 0.999 \\
\rowcolor{bnc!12} 0.07850 & 0.095692 & 0.095705 & 0.01299 & +0.500 & 0.999 \\
\rowcolor{bnc!12} 0.07875 & 0.095678 & 0.095692 & 0.01458 & +0.500 & 0.999 \\
\rowcolor{bnc!12} 0.07900 & 0.095663 & 0.095674 & 0.01199 & +0.500 & 0.999 \\
\rowcolor{bnc!12} 0.07925 & 0.095647 & 0.095661 & 0.01415 & +0.500 & 0.999 \\
\rowcolor{bnc!12} 0.07950 & 0.095631 & 0.095644 & 0.01340 & +0.500 & 0.999 \\
\rowcolor{bnc!12} 0.07975 & 0.095614 & 0.095627 & 0.01273 & +0.500 & 1.000 \\
\rowcolor{bnc!12} 0.08000 & 0.095597 & 0.095610 & 0.01314 & +0.500 & 0.999 \\
\rowcolor{bnc!12} 0.08025 & 0.095579 & 0.095592 & 0.01339 & +0.500 & 0.999 \\
\rowcolor{bnc!12} 0.08050 & 0.095560 & 0.095573 & 0.01327 & +0.500 & 1.000 \\
\rowcolor{bnc!12} 0.08075 & 0.095541 & 0.095554 & 0.01400 & +0.500 & 1.000 \\
\rowcolor{bnc!12} 0.08100 & 0.095521 & 0.095533 & 0.01326 & +0.500 & 1.000 \\
\rowcolor{bnc!12} 0.08125 & 0.095500 & 0.095512 & 0.01328 & +0.500 & 1.000 \\
\rowcolor{bnc!12} 0.08150 & 0.095478 & 0.095490 & 0.01239 & +0.500 & 1.000 \\
\rowcolor{bnc!12} 0.08175 & 0.095455 & 0.095469 & 0.01387 & +0.500 & 1.000 \\
\rowcolor{bnc!12} 0.08200 & 0.095432 & 0.095444 & 0.01273 & +0.500 & 1.000 \\
\rowcolor{bnc!12} 0.08225 & 0.095408 & 0.095420 & 0.01336 & +0.500 & 1.000 \\
\rowcolor{bnc!12} 0.08250 & 0.095382 & 0.095395 & 0.01306 & +0.500 & 1.000 \\
\rowcolor{bnc!12} 0.08275 & 0.095356 & 0.095368 & 0.01223 & +0.500 & 1.000 \\
\rowcolor{bnc!12} 0.08300 & 0.095329 & 0.095341 & 0.01317 & +0.500 & 1.000 \\
\rowcolor{bnc!12} 0.08325 & 0.095301 & 0.095313 & 0.01293 & +0.500 & 1.001 \\
\rowcolor{bnc!12} 0.08350 & 0.095271 & 0.095285 & 0.01478 & +0.500 & 1.001 \\
\rowcolor{bnc!12} 0.08375 & 0.095241 & 0.095253 & 0.01312 & +0.500 & 1.001 \\
\rowcolor{bnc!12} 0.08400 & 0.095209 & 0.095220 & 0.01189 & +0.500 & 1.001 \\
\rowcolor{bnc!12} 0.08425 & 0.095176 & 0.095189 & 0.01396 & +0.500 & 1.001 \\
\rowcolor{bnc!12} 0.08450 & 0.095142 & 0.095154 & 0.01321 & +0.500 & 1.002 \\
\rowcolor{bnc!12} 0.08475 & 0.095106 & 0.095119 & 0.01401 & +0.500 & 1.002 \\
\rowcolor{bnc!12} 0.08500 & 0.095069 & 0.095081 & 0.01285 & +0.500 & 1.002 \\
\rowcolor{bnc!12} 0.08525 & 0.095030 & 0.095044 & 0.01395 & +0.500 & 1.002 \\
\rowcolor{bnc!12} 0.08550 & 0.094990 & 0.095003 & 0.01311 & +0.500 & 1.003 \\
\rowcolor{bnc!12} 0.08575 & 0.094948 & 0.094960 & 0.01203 & +0.501 & 1.003 \\
\rowcolor{bnc!12} 0.08600 & 0.094905 & 0.094917 & 0.01300 & +0.501 & 1.004 \\
\rowcolor{bnc!12} 0.08625 & 0.094859 & 0.094870 & 0.01180 & +0.501 & 1.004 \\
\rowcolor{bnc!12} 0.08650 & 0.094812 & 0.094825 & 0.01413 & +0.502 & 1.005 \\
\rowcolor{bnc!12} 0.08675 & 0.094762 & 0.094773 & 0.01146 & +0.504 & 1.006 \\
\rowcolor{bnc!12} 0.08700 & 0.094710 & 0.094722 & 0.01251 & +0.505 & 1.007 \\
\rowcolor{bnc!12} 0.08725 & 0.094656 & 0.094669 & 0.01381 & +0.509 & 1.009 \\
\rowcolor{bnc!12} 0.08750 & 0.094599 & 0.094611 & 0.01302 & +0.511 & 1.010 \\
\rowcolor{bnc!12} 0.08775 & 0.094539 & 0.094552 & 0.01321 & +0.461 & 1.057 \\
\bottomrule
\end{NiceTabular}\hspace{1em}
\begin{NiceTabular}[baseline=t,colortbl-like]{cccccc}
\toprule
$T$ & $A_{\rm th}$ & $A$ & $\Delta$ & $b$ & $S/S_{\rm th}$ \\
\midrule
\rowcolor{sph!12} 0.08800 & 0.094286 & 0.094312 & 0.02804 & -0.461 & 1.033 \\
\rowcolor{sph!12} 0.08825 & 0.094019 & 0.094046 & 0.02963 & -0.104 & 1.064 \\
\rowcolor{sph!12} 0.08850 & 0.093753 & 0.093779 & 0.02722 & -0.071 & 1.046 \\
\rowcolor{sph!12} 0.08875 & 0.093489 & 0.093515 & 0.02785 & -0.062 & 1.032 \\
\rowcolor{sph!12} 0.08900 & 0.093226 & 0.093252 & 0.02765 & -0.056 & 1.030 \\
\rowcolor{sph!12} 0.08925 & 0.092965 & 0.092991 & 0.02762 & -0.037 & 1.024 \\
\rowcolor{sph!12} 0.08950 & 0.092705 & 0.092733 & 0.02938 & -0.036 & 1.025 \\
\rowcolor{sph!12} 0.08975 & 0.092447 & 0.092476 & 0.03059 & -0.033 & 1.021 \\
\rowcolor{sph!12} 0.09000 & 0.092190 & 0.092216 & 0.02752 & -0.028 & 1.019 \\
\rowcolor{sph!12} 0.09025 & 0.091935 & 0.091961 & 0.02846 & -0.026 & 1.018 \\
\rowcolor{sph!12} 0.09050 & 0.091681 & 0.091707 & 0.02808 & -0.025 & 1.017 \\
\rowcolor{sph!12} 0.09075 & 0.091429 & 0.091457 & 0.03109 & -0.024 & 1.016 \\
\rowcolor{sph!12} 0.09100 & 0.091177 & 0.091203 & 0.02824 & -0.022 & 1.015 \\
\rowcolor{sph!12} 0.09125 & 0.090928 & 0.090951 & 0.02578 & -0.014 & 1.016 \\
\rowcolor{sph!12} 0.09150 & 0.090679 & 0.090705 & 0.02810 & -0.013 & 1.013 \\
\rowcolor{sph!12} 0.09175 & 0.090432 & 0.090456 & 0.02618 & -0.013 & 1.012 \\
\rowcolor{sph!12} 0.09200 & 0.090186 & 0.090211 & 0.02776 & -0.013 & 1.011 \\
\rowcolor{sph!12} 0.09225 & 0.089942 & 0.089965 & 0.02606 & -0.011 & 1.009 \\
\rowcolor{sph!12} 0.09250 & 0.089699 & 0.089726 & 0.02986 & -0.008 & 1.009 \\
\rowcolor{sph!12} 0.09275 & 0.089457 & 0.089482 & 0.02786 & -0.011 & 1.009 \\
\rowcolor{sph!12} 0.09300 & 0.089217 & 0.089240 & 0.02683 & -0.011 & 1.009 \\
\rowcolor{sph!12} 0.09325 & 0.088977 & 0.089002 & 0.02713 & -0.008 & 1.007 \\
\rowcolor{sph!12} 0.09350 & 0.088739 & 0.088766 & 0.02965 & -0.007 & 1.007 \\
\rowcolor{sph!12} 0.09375 & 0.088503 & 0.088527 & 0.02702 & -0.007 & 1.007 \\
\rowcolor{sph!12} 0.09400 & 0.088267 & 0.088292 & 0.02803 & -0.007 & 1.006 \\
\rowcolor{sph!12} 0.09425 & 0.088033 & 0.088056 & 0.02550 & -0.006 & 1.006 \\
\rowcolor{sph!12} 0.09450 & 0.087800 & 0.087824 & 0.02727 & -0.006 & 1.005 \\
\rowcolor{sph!12} 0.09475 & 0.087569 & 0.087591 & 0.02500 & -0.006 & 1.005 \\
\rowcolor{sph!12} 0.09500 & 0.087338 & 0.087361 & 0.02642 & -0.005 & 1.005 \\
\rowcolor{sph!12} 0.09525 & 0.087109 & 0.087133 & 0.02737 & -0.005 & 1.004 \\
\rowcolor{sph!12} 0.09550 & 0.086881 & 0.086905 & 0.02783 & -0.004 & 1.004 \\
\rowcolor{sph!12} 0.09575 & 0.086654 & 0.086677 & 0.02623 & -0.005 & 1.004 \\
\rowcolor{sph!12} 0.09600 & 0.086429 & 0.086452 & 0.02667 & -0.004 & 1.003 \\
\rowcolor{sph!12} 0.09625 & 0.086204 & 0.086225 & 0.02402 & -0.004 & 1.003 \\
\rowcolor{sph!12} 0.09650 & 0.085981 & 0.086003 & 0.02623 & -0.003 & 1.003 \\
\rowcolor{sph!12} 0.09675 & 0.085759 & 0.085781 & 0.02573 & -0.003 & 1.003 \\
\rowcolor{sph!12} 0.09700 & 0.085538 & 0.085560 & 0.02594 & -0.003 & 1.003 \\
\rowcolor{sph!12} 0.09725 & 0.085318 & 0.085342 & 0.02899 & -0.003 & 1.002 \\
\rowcolor{sph!12} 0.09750 & 0.085099 & 0.085120 & 0.02508 & -0.002 & 1.002 \\
\rowcolor{sph!12} 0.09775 & 0.084881 & 0.084902 & 0.02462 & -0.002 & 1.002 \\
\rowcolor{sph!12} 0.09800 & 0.084665 & 0.084687 & 0.02593 & -0.002 & 1.002 \\
\rowcolor{sph!12} 0.09825 & 0.084449 & 0.084472 & 0.02644 & -0.002 & 1.002 \\
\rowcolor{sph!12} 0.09850 & 0.084235 & 0.084256 & 0.02477 & -0.002 & 1.002 \\
\rowcolor{sph!12} 0.09875 & 0.084022 & 0.084044 & 0.02613 & -0.002 & 1.001 \\
\rowcolor{sph!12} 0.09900 & 0.083809 & 0.083830 & 0.02463 & -0.002 & 1.001 \\
\rowcolor{sph!12} 0.09925 & 0.083598 & 0.083619 & 0.02522 & -0.001 & 1.001 \\
\rowcolor{sph!12} 0.09950 & 0.083388 & 0.083409 & 0.02509 & -0.001 & 1.001 \\
\rowcolor{sph!12} 0.09975 & 0.083179 & 0.083200 & 0.02512 & -0.001 & 1.001 \\
\rowcolor{sph!12} 0.10000 & 0.082971 & 0.082992 & 0.02523 & -0.001 & 1.001 \\
\bottomrule
\end{NiceTabular}
\caption{Quintic potential. The data underlying Figures.~\ref{fig:quinticAm} and~\ref{fig:quinticbS} (transition at $T_c\simeq 0.0878$). $A$, $b$, and $S/S_{\rm th}$ are the Borel--Pad\'e values and $\Delta=|A-A_{\rm th}|/A_{\rm th}$ (\%). 
%Rows are shaded by phase: \textcolor{bnc}{bounce} ($T<T_c$) and \textcolor{sph}{sphaleron} ($T>T_c$).
}
\label{tab:fvst-quintic}
\end{table}
\begin{figure}[tp!]
\centering
\begin{minipage}[c]{0.5\textwidth}
  \centering
  \includegraphics[width=\textwidth]{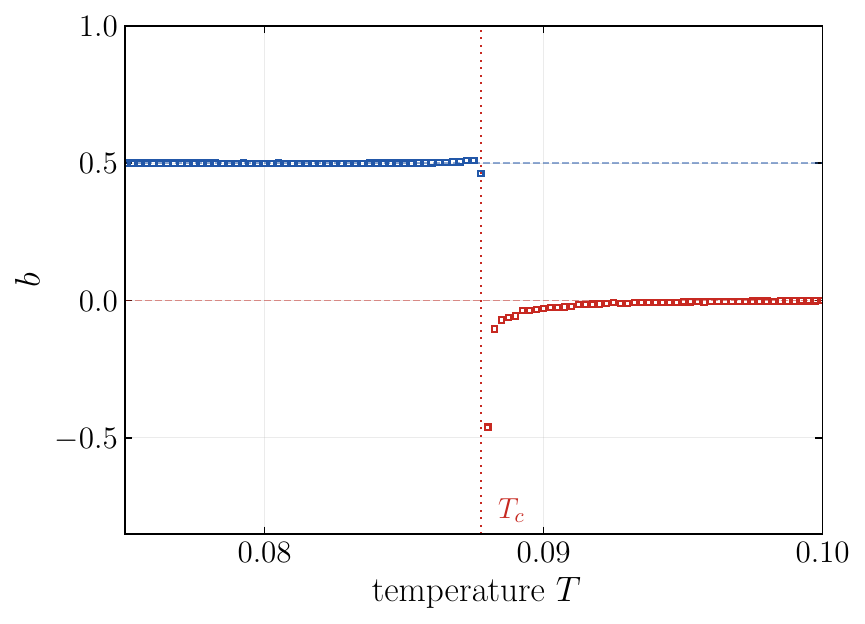}
\end{minipage}\hfill
\begin{minipage}[c]{0.5\textwidth}
  \centering
  \includegraphics[width=\textwidth]{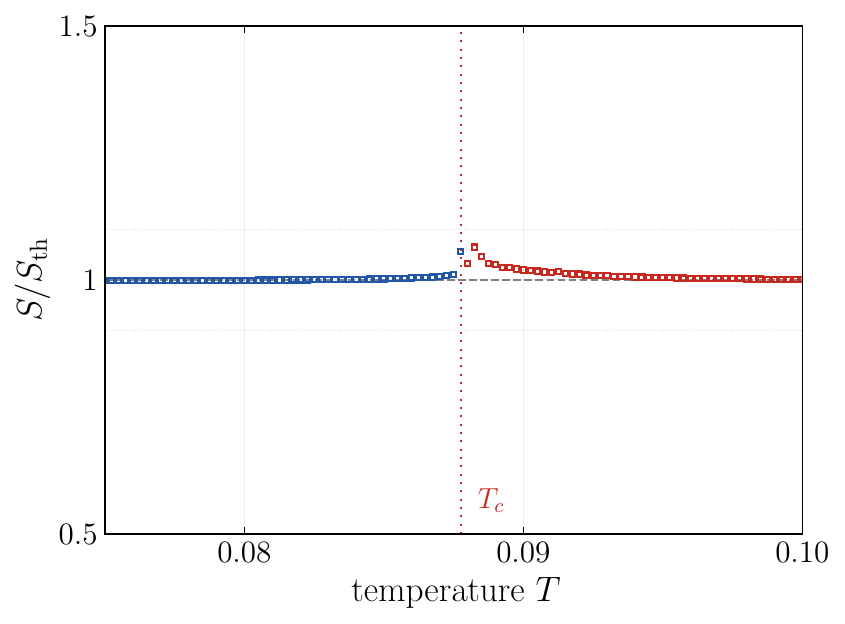}
\end{minipage}
\caption{Quintic potential. \emph{Left:} characteristic exponent $b(T)$ 
against the theoretical values $0$ and $1/2$ (dashed). \emph{Right:} Stokes ratio $S/S_{\rm th}$, near unity away from $T_{c}$ on both sides and deviating near $T_{c}$.}
\label{fig:quinticbS}
\end{figure}

The Borel singularities $A(T)$ are shown as open circles in Figure~\ref{fig:quinticAm}, lying on the periodic-bounce line $S_b(T)$ for $T<T_c$ and on the sphaleron line $\beta V_{\rm top}$ for $T>T_c$. 
The extracted $A(T)$ is listed in Table~\ref{tab:fvst-quintic}, using the perturbative free-energy series only. As for the cubic, the relative gap $\Delta$ is of order $10^{-2}\%$ across the window, rising to at most $\Delta\simeq 0.03\%$. In a narrow window around $T_c$, the two classical actions become nearly equal, so the two leading Borel singularities lie close together. The Borel--Pad\'e extraction $A(T)$ nonetheless follows the dominant action on both sides of $T_c$, including the crossing region, where it reproduces the kink that signals the first-order transition.

The left panel of Figure~\ref{fig:quinticbS} shows $b(T)$ taking the values $0$ on the hot sphaleron side and $1/2$ on the cold bounce side. These are the values expected from the zero modes, as in the cubic case. It departs from $0$ and $1/2$ only across $T_c$, where the two saddle contributions are both relevant.

The Stokes ratio $S/S_{\rm th}$ in the right panel of Figure~\ref{fig:quinticbS} lies close to unity away from $T_c$, so the extracted Stokes constant reproduces the Affleck prefactor \eqref{eq:Saff}. On the cold bounce side, the approach to unity from below is gradual. The deviations are highest at $T_c$, where the two leading Borel singularities are near-degenerate. We quote $b$ and $S$ only away from this region, and Table~\ref{tab:fvst-quintic} lists their values.

\section*{Acknowledgment}
S.D.H., K.L., and S.L. are supported by KIAS Grants PG096301, PG096201, and PG056502, respectively.

\newpage

% \bibliographystyle{IEEEtran}
% \bibliography{main}

\bibliographystyle{JHEP}
\bibliography{draft}

\end{document}